\documentstyle[twoside,fleqn,espcrc2]{article}
\input epsf
\newcommand{\eg}{{\em e.g.\ }}

\newcommand{\vs}{{\em vs.\ }}

\begin{document}
\title {
\begin{flushright}
{\normalsize IIT-HEP-95/7\\
\vspace {-.15 in}
hep-ex/9512002
}
\end{flushright}
\vskip 0.2in
\large Charm2000: A $>$10$^8$-charm experiment for the turn of the
millennium\thanks{Invited talk presented at the
Conference on Production and Decay of Hyperons, Charm and Beauty Hadrons,
Strasbourg, France,
5--8 September 1995.}}
\author{ Daniel M. Kaplan\thanks{E-mail: kaplan@fnal.gov}
     \\ {Illinois Institute of Technology, Chicago, IL 60616} %(institution)
	\\
         }  % end of \author

\begin{abstract}
I discuss the physics reach of a fixed-target charm experiment which can
reconstruct $>$10$^8$ charm decays, three orders of magnitude beyond the
largest extant sample. Such an experiment may run at Fermilab shortly after
the Year 2000. In addition to ``programmatic" charm physics such as
spectroscopy, lifetimes, and tests of QCD, this ``Charm2000" experiment will
have significant sensitivity to new physics in the areas of {\em CP}
violation, flavor-changing neutral-current and lepton-number-violating decays,
and mixing, and could observe direct {\em CP} violation in Cabibbo-suppressed
decays at the level predicted by the Standard Model.
\end{abstract}
\date{}
\maketitle
\addtocounter{footnote}{-2}

\section{Introduction}

Charm experiments have made important contributions to our effort to test the
Standard Model and search beyond it. I discuss in this paper the prospects for
increased contributions in the years ahead, and I argue for a new fixed-target
experiment to exploit to the full the demonstrated ability of the Fermilab
Tevatron to produce very large samples of charm decays which can be
reconstructed with small background.

\begin{figure}[htb]
\centerline{\epsfysize=2 in\epsffile{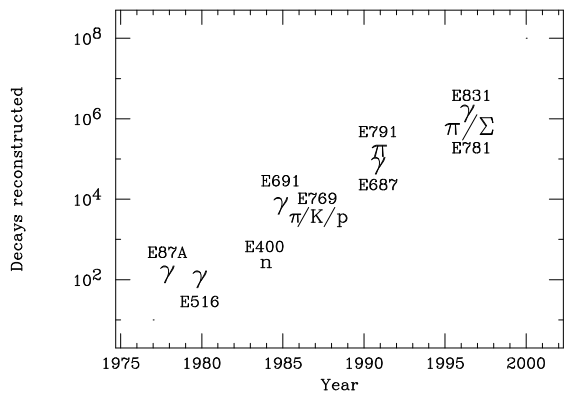}}
\vspace{-0.25in}
\caption{Yield of reconstructed charm \vs year of run
for completed or approved Fermilab fixed-target charm experiments with
the highest statistics for their generation; symbols indicate type of beam
employed.\label{history}}
\end{figure}

Following the more-or-less simultaneous discovery of the charm quark in
fixed-target~\cite{Ting} and $e^+e^-$ collisions~\cite{Richter}, for many
years experiments at $e^+e^-$ colliders dominated the study of charmed
particles. Starting in $\approx$1985, silicon vertex detectors made
fixed-target experiments once again competitive. More recently, advances in
data acquisition bandwidth and offline computing power have allowed the
recording of the very large unbiased event samples of Fermilab E769 and E791.
In parallel with the development of higher-intensity photon beams, these
developments have allowed exponential growth in the sensitivity of
fixed-target charm experiments, as illustrated in Fig.~\ref{history}.
Current charm samples are in the $10^5$-reconstructed-decays range, the
Fermilab hadroproduction experiment E791 having the largest sample at
$\approx$250,000 events~\cite{Wiss}. Fermilab experiments
E831 and E781 aim to reach the $10^6$-event level in the 1996/7 Fermilab
fixed-target run.

At the CHARM2000 Workshop~\cite{CHARM2000} the prospect of pushing to
substantially higher sensitivity was considered. A sample of $>$10$^8$ events
was identified as a desirable goal for a next-generation experiment. Such
sensitivity could bring within reach the observation of {\em CP} violation in
charm decay at the level expected in the Standard Model, while extending
sensitivity to new physics by two orders of magnitude in statistical power.
As described below, such an advance appears feasible in an experiment to run in
the Year $\approx$2000.

\section{High-Impact Charm Physics}

``High-impact" denotes measurements which are particularly sensitive to new,
non-Standard-Model physics~\cite{Bigi89}.  The Standard Model (SM) contains two
key  mysteries: the origin of mass and the existence of multiple fermion
generations.
While the former mystery may be resolved
by the LHC, the latter appears to originate at
higher mass scales, which can only
be studied indirectly. Such effects as {\em CP}  violation, mixing, and
flavor-changing neutral or lepton-number-violating currents may hold the key to
physics at these new scales~\cite{Hewett,Pakvasa,Sokoloff}.
Because in the charm
sector the SM contributions to these effects are small, these are areas in
which charm studies can provide unique information.
In contrast, in the $s$- and $b$-quark sectors in which such studies are
typically pursued, there are large SM contributions to mixing and {\em CP}
violation~\cite{Rosner}, which for new-physics searches constitute backgrounds.

Table~\ref{tab:sens} summarizes current sensitivities in high-impact charm
physics.
Also indicated is the sensitivity achievable in Charm2000, based on yield
estimates such as those in Table~\ref{tab:yields}.
No other proposed experiment is competitive in
reach~\cite{Kaplan-TCF}.

I next discuss each physics topic
in more detail,  then summarize the salient aspects of the
Charm2000  experiment.

\subsection{Direct {\em CP} violation}

The Standard Model predicts direct {\em CP} violation at the $\sim$10$^{-3}$
level in singly-Cabibbo-suppressed decays (SCSD) of
charm~\cite{Bigi89,Golden,Pugliese,Burdman}. {\em CP} violation  in
Cabibbo-favored (CFD) or doubly-Cabibbo-suppressed (DCSD) modes would
be a clear signature of new physics~\cite{Burdman,Bigi94}. Asymmetries in all
three  categories could reach $\sim$10$^{-2}$ in such scenarios as non-minimal
supersymmetry~\cite{Bigi94} and in left-right-symmetric
models~\cite{Pakvasa,Yaouanc}. There are also expected SM asymmetries of
$\approx\!3.3\times10^{-3}$ ($=2\,Re(\epsilon_K)$) due to $K^0$ mixing in such
modes\footnote{To avoid such cumbersome notations as
$D^0 (\overline {D^0})\to K^\mp\pi^\pm$, here and elsewhere in
this paper charge-conjugate states are generally implied even when not
stated.}  as $D^+\to K_S\pi^+$ and $K_S \ell\nu$~\cite{Xing}, which should be
observed in Charm2000 or even in predecessor
experiments. While $K^0$-induced {\em CP} asymmetries might
teach us little we don't already know,
they will at least constitute a calibration
for the experimental systematics of  asymmetries at the $10^{-3}$ level.
However, Bigi has pointed out that a small new-physics contribution to the DCSD
rate could amplify these asymmetries to $\cal{O}$$(10^{-2})$~\cite{Bigi94}.

The signal for direct {\em CP} violation is an absolute rate difference between
decays of particle and antiparticle to charge-conjugate final states $f$ and
${\bar f}$:
\begin{equation}
\label{eq:CP}
A=\frac{\Gamma(D\to f)-\Gamma({\overline D}\to{\bar f})}
{\Gamma(D\to f)+\Gamma({\overline D}\to{\bar f})}\,.
\end{equation}
Extrapolation from sensitivity in E687~\cite{Frabetti} implies {\em CP}
sensitivities in Charm2000
in SCSD modes of $\approx$10$^{-3}$ at 90\% confidence~\cite{Kaplan-TCF}.
Because of the $D\overline D$ production asymmetry, in
fixed-target experiments the rates in Eq.~\ref{eq:CP} are in
practice normalized to the observed rates in Cabibbo-favored modes. The
ratiometric nature of the measurement reduces sensitivity to systematic biases,
but at the $10^{-3}$ level systematics will need to be studied carefully.

Since one CFD mode must be used for normalization, the search for direct
{\em CP} violation in CFD modes is actually a search for {\em CP}-asymmetry
{\em differences} among various modes. Given the differing final-state
interactions, if new physics causes {\em CP} violation in CFD modes,
such differences are not unlikely.
The yields indicated in Table~\ref{tab:yields} imply
{\em CP} sensitivity at the few\,$\times10^{-4}$ level
in Charm2000 for $D^0\to K^- \pi^+\pi^-\pi^+$, normalized to the
production asymmetry observed in $D^0\to K^- \pi^+$.
For DCSD modes, extrapolations from preliminary E791
results on $D^+\to K^+\pi^+\pi^-$~\cite{Purohit-Weiner}
and CLEO's observation of $D^0\to K^+\pi^-$~\cite{Cinabro}
suggest {\em CP} sensitivity
in Charm2000 at the few$\times10^{-3}$ to $\approx$10$^{-2}$
level~\cite{Kaplan-TCF}.
These extrapolations are conservative and ignore expected  improvements in
vertex resolution and particle identification. Detailed simulations  are
underway to assess these effects.

SM predictions for direct {\em CP} violation are rather uncertain, since they
require assumptions for final-state phase shifts as well as CKM matrix
elements~\cite{Burdman,Bigi94}; the predictions given in Table~\ref{tab:sens}
are representative, but the theoretical uncertainties are probably larger than
indicated there~\cite{Buccella2}. However, given the order of magnitude
expected in charm decay, the Charm2000 experiment might make the first
observation of direct {\em CP} violation outside the strange sector, or indeed
the first observation anywhere if (as may well be the case~\cite{Paschos,Lu})
signals prove too small for detection in the next
round of $K^0$~\cite{KTeV,NA48} and hyperon~\cite{E871}
experiments~\cite{Kaplan-4seas}.

\subsection{Flavor-Changing Neutral Currents}

Charm-changing neutral currents are forbidden at tree level in the Standard
Model due to the GIM mechanism~\cite{GIM}. They can proceed via loops at rates
which are predicted to be unobservably small, \eg for $D^0\to\mu^+\mu^-$
(which suffers also from helicity suppression in the SM) the predicted
branching ratio is $\sim10^{-19}$~\cite{Gorn,Pakvasa,Hewett}, and for
$D^+\to\pi^+\mu^+\mu^-$ it is $\sim10^{-10}$~\cite{Babu,Hewett}. Long-distance
effects increase these predictions by some orders of magnitude, but they
remain of order $10^{-15}$ to $10^{-8}$~\cite{Pakvasa,Schwartz,Burdmanetal}.
Various extensions of the SM~\cite{Babu,Lepto} predict effects substantially
larger than this, for example in models with a fourth generation, both
$B(D^+\to\pi^+\mu^+\mu^-)$ and $B(D^0\to\mu^+\mu^-)$ can be as large as
$10^{-9}$~\cite{Babu}. Experimental sensitivities are now in the range
$\sim10^{-4}$ to $10^{-5}$~\cite{Aitala,PDG,WA92,Sheldon,E653}
and are expected to reach $\sim10^{-5}$ to $10^{-6}$ in E831~\cite{Cumalat}.

While Charm2000 aims at a single-event branching-ratio sensitivity
of $\approx\!10^{-9}$, FCNC limits are typically background-limited, so
sensitivites can be expected to improve as the square root of the number
of events reconstructed. In some cases, however, more dramatic improvement may
result from
improved lepton identification. For $D^+\to\pi^+\mu^+\mu^-$, scaling
E791 sensitivity~\cite{Aitala} by a factor of $\sqrt{2000}$ gives
few$\times10^{-7}$ sensitivity in Charm2000. This
estimate may
be conservative, since the simple muon detection scheme employed by E791 (one
layer of scintillation counters following 2.5\,m of steel equivalent) resulted
in a (momentum-dependent)  $\pi$-$\mu$ misidentification probability ranging
from 4.5 to 20\%~\cite{Aitala},
and it should be possible to reduce this to $\approx$1\% in
Charm2000. With modern calorimetry for electron identification one expects to
do almost  as well for $\pi e e$ as for $\pi\mu\mu$. For $D^0\to\mu^+\mu^-$ and
$e^+e^-$,  extrapolation from WA92~\cite{WA92}
implies sensitivity of $10^{-7}$ per mode.

\subsection{Lepton-Number-Violating Decays}

There are two lepton-number-violating effects which can be sought:
decays violating conservation of lepton number (LNV) and decays violating
conservation of lepton-family number (LFNV). LFNV decays (such as
$D^0\to\mu^\pm e^\mp$) are expected in theories with leptoquarks~\cite{Lepto},
heavy neutrinos~\cite{Hewett}, extended technicolor~\cite{Techni}, etc.
LNV decays (such as $D^+\to K^- e^+ e^+$ or $\Sigma^+\pi^+e^-$)
can arise in GUTs and have been
postulated to play a role in the development of the baryon asymmetry of the
Universe~\cite{cosmic-baryon}. Since no known fundamental principle forbids
either type of decay, it is of interest to search for
them as sensitively as possible.

Although much smaller decay widths can be probed in $K$ decays, there are
simple theoretical arguments why LFNV charm decays are nevertheless
worth seeking.
For example, if these effects arise through Higgs
exchange, whose couplings are proportional to mass, they will couple more
strongly to charm than to strangeness~\cite{Bigi87}.
Furthermore, LFNV currents
may couple to up-type quarks more strongly than to
down-type~\cite{Lepto,Hadeed}.

As shown in Table~\ref{tab:sens}, the best existing limits come in most cases
from the $e^+e^-$ experiments Mark II, ARGUS, and CLEO (although the
hadroproduction experiment Fermilab E653 dominates in modes with same-sign
dimuons) and are typically at the $10^{-3}-10^{-4}$ level~\cite{Sheldon,E653}.
E831 expects to lower these limits to $\sim10^{-6}$~\cite{Cumalat},
and Charm2000 should reach $\sim10^{-7}$.

\subsection{Mixing and Indirect {\em CP} Violation}

$ D^0\overline {D^0} $ mixing may be one of the more promising places to
look for low-energy manifestations of physics beyond the Standard Model.
SM contributions to $|\Delta M_D|$ are estimated~\cite{Burdman,mixing} to give
$r_{\rm mix}
\sim (\Delta M_D/\Gamma_D)^2<10^{-8}$; any observation at a
substantially higher level will be clear evidence of new
physics.\footnote{Earlier estimates~\cite{Wolfenstein_mixing}
that long-distance effects can give $\Delta M_D/\Gamma_D \sim 10^{-2}$
are claimed to have been disproved~\cite{Burdman}, though there
 remain skeptics~\cite{Bigi94,Wolfenstein}.}
Many nonstandard models predict much larger effects.
An interesting example  is the multiple-Higgs-doublet model
lately expounded by Hall and Weinberg~\cite{Hall-Weinberg}, in  which
$|\Delta M_D|$ can be as large as $10^{-4}$\,eV, approaching the current
experimental limit. In this model $K^0$ {\em CP} violation arises from the
Higgs sector, and {\em CP} violation in the beauty sector is expected to be
small, which
emphasizes the importance of exploring rare phenomena in {\em
all} quark sectors. The large mixing contribution arises from flavor-changing
neutral-Higgs exchange (FCNE)~\cite{Wu},
which can be constrained to satisfy the GIM
mechanism for $K^0$ decay by assuming small phase factors
($\sim10^{-3}$).\footnote{This is in distinction to the original ``Weinberg
model" of {\em CP}
violation~\cite{Weinberg}, in which FCNE was suppressed by assuming a discrete
symmetry such that one Higgs gave mass to up-type quarks and another to
down-type.} Many other authors have also considered multiple-Higgs effects in
charm mixing~\cite{Hadeed,Datta,Bigi-Sanda,multiple-Higgs,Bigietal}.
Large mixing in charm can also arise in theories with
supersymmetry~\cite{Datta,Nir}, technicolor~\cite{Techni},
leptoquarks~\cite{Lepto}, left-right symmetry~\cite{seesawLR},
or a fourth generation~\cite{Pakvasa,Babu}.

The experimental situation regarding $D^0\overline {D^0}$ mixing is complicated
by the presence of DCSD. Since both effects can lead to the same final states,
one needs to distinguish them using time-resolved measurements~\cite{Bigi87}.
In the notation of Refs.~\cite{Blaylock} and \cite{Browder}, the time
dependence for wrong-sign decay is given by
$$\Gamma(D^0(t)\to K^+\pi^-) = |B|^2|\frac{q}{p}|^2~\times \qquad\qquad\qquad$$
\vspace{-0.2 in}
\begin{eqnarray}
\label{eq:mixing}
\frac{e^{-\Gamma t}}{4}
 \{4 |\lambda|^2
+ (\Delta M^2 + \frac{\Delta\Gamma^2}{4})t^2 + \nonumber \\
2{\rm Re}(\lambda)\Delta\Gamma t + 4 {\rm Im}(\lambda)\Delta Mt\}\,,
\end{eqnarray}
and there is a similar expression for $\overline{D^0}\to K^-\pi^+$
in which $\lambda$ is replaced by ${\bar \lambda}$.
In Eq.~\ref{eq:mixing} the first term on the right-hand side is the DCSD
contribution, which peaks at $t=0$;
the second is the mixing contribution, which peaks at 2 $D^0$
lifetimes because of  the factor $t^2$; and the third and fourth
terms reflect interference between mixing and DCSD and peak at 1 lifetime due
to the factor $t$.
$\lambda$ and $\bar \lambda$ can acquire nonzero phases through indirect {\em
CP} violation or through final-state interactions~\cite{Browder,Wolfenstein}.
For small values of $r_{\rm mix}$ experimental sensitivity to mixing can
be enhanced by interference~\cite{Liu}. However, at present levels of
sensitivity, allowing an arbitrary interference phase when fitting decay-time
distributions reduces the stringency of the resulting
limit~\cite{E691,Purohit95}.

Extrapolation by $\sqrt{2000}$ from preliminary E791 results~\cite{Purohit95}
suggests sensitivity of
$\approx$2$\times10^{-5}$ in Charm2000 (neglecting interference), which with
improvements in particle identification and resolution for the tagging pion
might approach $10^{-5}$.
Since the interference term is linear in $\Delta M_D$ while the mixing term is
quadratic, the ratio of the interference and mixing contributions
goes as $1/\Delta M_D$. Thus as experimental sensitivity
improves and smaller and smaller values of $\Delta M_D$ are probed,
interference becomes relatively more important. In a (model-dependent)
estimate of Charm2000 sensitivity based on the
prescription of Browder and Pakvasa~\cite{Browder},
the interference term improves sensitivity slightly,
and $10^{-5}$ sensitivity is obtained~\cite{Kaplan-TCF}.

Semileptonic decays offer a way to study mixing free from the effects of DCSD.
A preliminary result from E791 using $D^*$-tagged $D^0\to K e\nu$ events
indicates sensitivity at the $\approx$0.5\% level~\cite{Tripathi}.
Extrapolation by $\sqrt{2000}$ suggests $10^{-4}$ sensitivity in
Charm2000, but use of muonic decays as well, plus improvements in lepton
identification and resolution for the tagging pion,
may give significantly better sensitivity.
At the CHARM2000 Workshop, Morrison suggested $10^{-5}$ sensitivity may be
possible~\cite{Morrison}.

Liu has stressed the importance of setting limits on $\Delta\Gamma$ as well as
on $\Delta M$. Although typical extensions of the SM which predict large
$\Delta M$ also predict $\Delta M\gg\Delta\Gamma$~\cite{Blaylock,Browder},
from an experimentalist's viewpoint both should be measured if possible.
This can be done quite straightforwardly by comparing the lifetime measured for
{\em CP}-even modes (such as $K^+ K^-,\pi^+\pi^-$) with that for {\em CP}-odd
modes or (more simply) with modes of mixed {\em CP} (such as $K^-\pi^+$).
Liu has estimated the Charm2000 sensitivity at $\sim 10^{-5}-10^{-6}$
in $y^2\equiv (\Delta\Gamma/2\Gamma)^2$~\cite{Liu}.

\subsubsection{Indirect {\em CP} violation}
\label{indirect}

In the SM $D^0\overline {D^0}$ mixing is negligible, and any indirect {\em
CP}-violating asymmetries are expected to  be less than
$10^{-4}$~\cite{Bigi94}.
However, possible mixing signals at the $\approx$1\% level have been
reported~\cite{Cinabro,MkII-mixing}. Given the E691 mixing limit these
presumably
represent enhanced DCSD signals. If a significant portion of this rate
is mixing, new physics must be responsible~\cite{Burdman,Wolfenstein}.
Indirect {\em CP} violation at the $_\sim$\llap{$^<$}1\%
level is then possible~\cite{Bigi-Sanda,Bigi2000,Bigi94,Wolfenstein}.
Several authors have suggested that the {\em CP}-violating signal, which
arises from the interference term of Eq.~\ref{eq:mixing}, may be
easier to detect than the mixing
itself~\cite{Blaylock,Browder,Wolfenstein,Liu}. In particular, Browder and
Pakvasa~\cite{Browder} point out that in the difference
$\Gamma(D^0\to K^+\pi^-)-\Gamma(\overline{D^0} \to K^-\pi^+)$, the DCSD
and mixing components cancel, leaving only the fourth term of
Eq.~\ref{eq:mixing}. Thus if indirect {\em CP} violation is
appreciable, this is a particularly clear way to isolate the
interference term.

\section{Testing the Standard Model with Charm}

In addition to searches for effects due to new physics,
high-sensitivity charm measurements
address a variety of Standard-Model issues.
These have been discussed recently by Sokoloff~\cite{Sokoloff-Pisa}, Sokoloff
and Kaplan~\cite{Sokoloff}, and Wiss~\cite{Wiss}.

\subsection{Testing the heavy-quark effective theory}

Heavy-quark symmetry can be used to predict many nonperturbative properties of
hadrons containing a heavy quark
(including form factors as discussed below).
As a rigorous limit of QCD, HQET needs to be tested in its own right, but it is
also important as a method for
extracting $ V_{ub} $ and $ V_{cb} $ from $B$-decay measurements.
HQET can be tested in the charm sector
through its predictions~\cite{Wisgur2,Wisgur3,EHQ} for the masses and widths of
the orbitally-excited $ D^{**} $ mesons~\cite{Sokoloff}.
Charm2000 should achieve few-percent fractional errors
on the masses and widths of many $ D^{**} $ states, where present measurements
are at the $\approx$50\% level~\cite{E691,ARGUS,CLEO,E687}.
By probing the importance of finite-mass effects at the charm-quark mass,
such measurements will help establish to what extent HQET is applicable to
beauty~\cite{Rosner2000}.

\subsection{Semileptonic form factors}

Semileptonic form factors are a testing ground for
nonperturbative QCD effects~\cite{Wiss}.
They are also important for extraction of CKM matrix elements
from charm decay and for
{\it CP}-violation studies in beauty decay. For example,
the method proposed by Dunietz~\cite{Dunietz} for
measuring the unitarity-triangle angle $  \gamma $
using branching ratios for $B_d\to K^*\psi$ and $\rho^0 \psi$
requires knowledge of semileptonic
form factors and helicity amplitudes.
These should be the same in $ D $ as in $B $ decay, thus
precise measurements in the charm sector will be an important input.
Modeling the
$ D^+ \rightarrow K^{*0} \ell \nu $ and $ \rho^0 \ell \nu $
form factors with single-pole forms, the pole mass
should be measurable in Charm2000 to better than 1\%.
The polarization of the $ K^* $ (the ratio of longitudinal to transverse
form-factors) should be measurable with $\approx$percent statistical and
systematic uncertainties, and that of the $ \rho^0 $ with  few-percent
statistical accuracy.
$ D_S \rightarrow \phi l \nu $ should  be measured with similar precision,
providing another test of heavy-quark symmetry~\cite{Rosner2000}.

\subsection{Studying the CKM matrix with semileptonic decays}

Semileptonic decays can be used to measure
the CKM-matrix elements $ V_{cs} $ and $ V_{cd} $.
Currently, $| V_{cd}| $ and $ | V_{cs} |$ are known to $ \pm 5 \% $ and $ \pm
15 \% $ respectively~\cite{Rosner}.
{}From the branching ratios for the semileptonic decays
$ D^0 \rightarrow \pi^- l^+ \nu_{l} $
and $ D^0 \rightarrow K^- l^+ \nu_{l} $,
the ratio $ |
V_{cd} | / | V_{cs} | $ should be determined in Charm2000
with a statistical accuracy of $\sim$10$^{-3}$.

\subsection{Hadronic decays}

Hadronic decays of mesons containing heavy quarks have many interesting
applications. As noted above, they can be used to search for
direct {\em CP} asymmetries, the size of which depends on final-state phase
shifts. The phase shifts can be studied with branching-ratio and
Dalitz-plot analyses~\cite{Wiss}. Such studies test nonperturbative QCD models
and are relevant to direct {\em CP} violation in beauty~\cite{Rosner2000} and
charm~\cite{Pugliese,Buccella2} decays.
The resonant substructure in charm decay to multiparticle final states can also
be a QCD laboratory, with the possibility of clarifying the questions of
existence of glueballs and gluonic hybrids~\cite{Lipkin,Close-Lipkin}.

\section{A Next-Generation Charm Spectrometer}

\begin{figure}
\centerline{\epsfysize = 1.02 in \epsffile {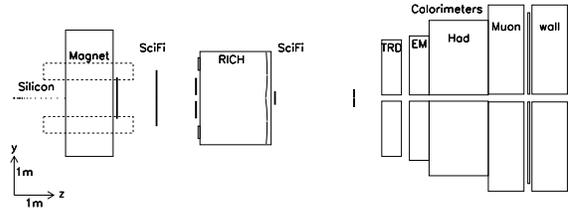}}
\caption [Spectrometer layout (bend view).]%
{Spectrometer layout (bend view).}
\label{app865}
\end{figure}

\begin{figure}[htb]
\vspace{0.1in}
\hspace{-.07in}\centerline{\epsfysize =1 in \epsffile {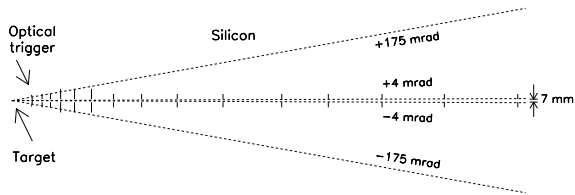}}
\caption [Detail of vertex region (showing optional optical impact-parameter
trigger).]%
{Detail of vertex region (showing optional optical impact-parameter
trigger).}
\label{detail865}
\end{figure}

A proposal is under development for a new
Fermilab experiment  to reconstruct $\approx$4$\times10^8$ charm decays,
$\approx$2000 times the largest extant charm sample, in the Year-$\approx$2000
fixed-target run.
The spectrometer
(Figs.~\ref{app865},~\ref{detail865})
is planned to be compact and of moderate cost
(\eg substantially cheaper than HERA-$B$~\cite{HERA-B}), but
with large acceptance, good resolution, and high-rate
tracking and particle identification. Tracking is done exclusively with
silicon or diamond~\cite{Tesarek} and
scintillating-fiber~\cite{Ruchti} detectors, allowing operation at a 5\,MHz
interaction rate. A
fast ring-imaging Cherenkov counter~\cite{Bari}
provides hadron identification, and
calorimeters (possibly augmented by a TRD)
identify electrons and allow first-level triggering on transverse
energy.
Triggering efficiently on charm while maintaining high livetime and a
manageable data rate to tape ($_\sim$\llap{$^<$}100\,MB/s)
is a significant challenge, requiring
hardware decay-vertex triggers~\cite{triggers}; first-level ``optical" triggers
may play a significant role~\cite{optrig,mul-jump}.\footnote{While
HERA-$B$ could be competitive with Charm2000 as a
charm experiment, it lacks the capabilities to trigger efficiently on charm
and to acquire the needed large data sample, and it probably has significantly
poorer vertex resolution as well.}
(More detailed discussions may be found in \cite{Kaplan2000} and
\cite{Kaplan95}.)

\section{Yield}
\label{sec:yields}

In 800$\,$GeV proton collisions with a high-$A$ target, charm is produced at a
rate of $\approx$7$\times10^{-3}$ per interaction~\cite{sigmaD}.
Thus at a 5$\,$MHz interaction rate in a
typical fixed-target run of $3\times10^6$ live beam seconds, $10^{11}$ charmed
particles are produced. The reconstructed-event yields in
representative modes are estimated
in Table~\ref{tab:yields}, with efficiencies derated for all-hadronic
modes under the assumption that the optical trigger described in \cite{P865} is
used for those modes.
For leptonic modes, the first-level
trigger rate should be sufficiently low to be
recorded directly.
The second half of the table gives yields extrapolated by a factor of 2000
from E791.
The total reconstructed sample is
well in excess of $10^8$ events. Given the factor $\approx$2
mass-resolution improvement
compared to E791, one can infer a factor $\approx$50 improvement in
statistical significance for typical decay modes.
No other proposed experiment is competitive in
reach.\footnote{The CHEOPS Letter of Intent to CERN~\cite{CHEOPS} and the
proposed Tau/Charm Factory~\cite{Toki} both aim at $\sim$10$^7$ reconstructed
charm.}

\section{Conclusions}

A fixed-target hadroproduction experiment (Charm2000)
capable of reconstructing $>$10$^8$ charm decays is feasible using detector,
trigger, and data
acquisition technologies which exist or are under development.
A typical factor $\approx50$ in statistical
significance of signals may be expected compared to E791,
possibly bringing within reach the observation of
Standard-Model {\em CP} violation in charm decay,
and extending searches for new physics by two orders of magnitude in
statistical power.

\section*{Acknowledgements}

I thank A. Fridman for the invitation to participate, as well
as for organizing so stimulating  a conference in such memorable surroundings.

{\footnotesize
\begin{table*}
\caption{Sensitivity to high-impact charm physics. \label{tab:sens}}
%\vspace{0.1in}
\begin{center}
\begin{tabular}{|l|l|l|l|l|}
\hline
& & Charm2000 & SM \\
\raisebox{1.5ex}[0pt]{Topic} & \raisebox{1.5ex}[0pt]{Limit$^*$} & Reach$^*$
& prediction \\
\hline\hline
Direct {\em CP} Viol.\
& & & \\ \hline
{}~$D^0\to K^- \pi^+$ & -0.009$<$$A$$<$0.027~\cite{Bartelt} &  & $\approx0$
(CFD)
\\
{}~$D^0\to K^- \pi^+\pi^-\pi^+$ & &  few$\,\times10^{-4}$ & $\approx0$ (CFD)
\\
{}~$D^0\to K^+ \pi^-$ &
& $10^{-3}-10^{-2}$ & $\approx0$ (DCSD) \\
{}~$D^+\to K^+ \pi^+ \pi^-$ &
& few$\,\times10^{-3}$ & $\approx0$ (DCSD) \\
{}~$D^0\to K^- K^+$ &
-0.11$<$$A$$<$0.16~\cite{Frabetti} & $10^{-3}$ & \\
& -0.028$<$$A$$<$0.166~\cite{Bartelt} & & \\
{}~$D^+\to K^- K^+\pi^+$ & -0.14$<$$A$$<$0.081~\cite{Frabetti}
& $10^{-3}$ & \\
{}~$D^+\to \overline {K^{*0}}K^+$
& -0.33$<$$A$$<$0.094~\cite{Frabetti} & $10^{-3}$ & $(2.8\!\pm\!0.8)\times
10^{-3}$~\cite{Pugliese} \\
{}~$D^+\to \phi\pi^+$ & -0.075$<$$A$$<$0.21~\cite{Frabetti}
& $10^{-3}$ & \\
{}~$D^+\to \eta\pi^+$ & & & $(-1.5\!\pm\!0.4)\times10^{-3}$~\cite{Pugliese}\\
{}~$D^+\to K_S\pi^+$ & & few$\times10^{-4}$ & $3.3\times10^{-3}$~\cite{Xing} \\
\hline
FCNC
& & & \\ \hline
{}~$D^0\to\mu^+\mu^-$ & $7.6\times 10^{-6}$~\cite{WA92} &$10^{-7}$
& $<3\times10^{-15}$~\cite{Hewett} \\
{}~$D^0\to \pi^0\mu^+\mu^-$ & $1.7\times10^{-4}$~\cite{E653} & $10^{-6}$ & \\
{}~$D^0\to \overline {K^0} e^+e^-$ & $17.0\times10^{-4}~\cite{Sheldon}$
& $10^{-6}$ & $<2\times10^{-15}$~\cite{Hewett} \\
{}~$D^0\to\overline {K^0}\mu^+\mu^-$ & $2.5\times10^{-4}$~\cite{E653}
& $10^{-6}$ & $<2\times10^{-15}$~\cite{Hewett} \\
{}~$D^+\to \pi^+e^+e^-$ & $6.6\times10^{-5}$~\cite{Aitala}
& few$\,\times10^{-7}$ & $<10^{-8}$~\cite{Hewett} \\
{}~$D^+\to \pi^+\mu^+\mu^-$ & $1.8\times10^{-5}$~\cite{Aitala}
& few$\,\times10^{-7}$ & $<10^{-8}$~\cite{Hewett} \\
{}~$D^+\to K^+ e^+e^-$ & $4.8\times10^{-3}$~\cite{Sheldon}
& few$\,\times10^{-7}$ & $<10^{-15}$~\cite{Hewett} \\
{}~$D^+\to K^+ \mu^+\mu^-$ & $8.5\times10^{-5}$~\cite{PDG}
& few$\,\times10^{-7}$ & $<10^{-15}$~\cite{Hewett} \\
{}~$D\to X_u+\gamma$ & & & $\sim10^{-5}$~\cite{Hewett} \\
{}~$D^0\to \rho^0\gamma$ & $1.4\times10^{-4}$~\cite{Hewett} & &
$(1-5)\times10^{-6}$~\cite{Hewett} \\
{}~$D^0\to \phi\gamma$ & $2\times10^{-4}$~\cite{Hewett} &
& $(0.1-3.4)\times10^{-5}$~\cite{Hewett} \\
\hline
LF or LN Viol.\
& & & \\ \hline
{}~$D^0\to\mu^\pm e^\mp$ & $1.0\times 10^{-4}$~\cite{PDG} & $10^{-7}$ & 0 \\
{}~$D^+\to\pi^+\mu^\pm e^\mp$ & $3.3\times 10^{-3}$~\cite{Sheldon}
& few$\times10^{-7}$ & 0 \\
{}~$D^+\to K^+ \mu^\pm e^\mp$ & $3.4\times 10^{-3}$~\cite{Sheldon}
& few$\times10^{-7}$ & 0 \\
{}~$D^+\to \pi^- \mu^+\mu^+$ & $2.2\times 10^{-4}$~\cite{E653}
& few$\times10^{-7}$ & 0 \\
{}~$D^+\to K^- \mu^+\mu^+$ & $3.3\times 10^{-4}$~\cite{E653}
& few$\times10^{-7}$ & 0 \\
{}~$D^+\to \rho^- \mu^+\mu^+$ & $5.8\times 10^{-4}$~\cite{E653}
& few$\times10^{-7}$ & 0 \\
\hline
Mixing
& & & \\ \hline
{}~${}^{^{(}}{\overline {D^0}}{}^{^{)}}\to K^\mp\pi^\pm$ &
$r<0.0037~\cite{E691},$ & $r<10^{-5},$ & \\
& $\Delta M_D<1.3\!\times\!10^{-4}$\,eV & $\Delta M_D<10^{-5}\,$eV &
$10^{-7}$\,eV~\cite{Burdman} \\
{}~${}^{^{(}}{\overline {D^0}}{}^{^{)}}\to K\ell\nu$ & & $r<10^{-5}$ & \\
\hline
\end{tabular}
\end{center}
$^*$at 90\% confidence level
\end{table*}
}

\begin{table*}
\footnotesize
\caption
{ Estimated yields of reconstructed decays (antiparticles
included) in Charm2000.}
\label{tab:yields}
\begin{center}
\begin{tabular}{|l|c|c|c|c|}
\hline
& & & acceptance & \\
\raisebox{1.5ex}[0pt]{mode} & \raisebox{1.5ex}[0pt]{charm frac.} &
\raisebox{1.5ex}[0pt]{BR (\%)}
& $\times$ efficiency & \raisebox{1.5ex}[0pt]{Charm2000 yield} \\
\hline
\hline
$D^0\to K^-\pi^+$ & 0.5 & 4.0 & $0.6\times0.1$ & $1.3\times10^8$ \\
$D^+\to K^{*0}\mu\nu$ &&&&\\
\qquad$\to K\pi\mu\nu$ & 0.25 & 2.7 & $0.4\times0.25$ & $7\times10^7$\\
all & 1 & $\approx0.1$ & $\approx 0.4\times0.1$ &
$\approx4\times10^8$\\
\hline
\hline
& analysis$^*$ & & E791 yield$^*$ & \\
\hline
\hline
$D^+\to K^-\pi^+\pi^+$ & FCNC & 9.1 & $37000\pm200$ & $(7\pm0.001)\times10^7$
\\
$D^+\to K_S\pi^+$ & & 0.94 & & $(7\pm0.003)\times10^6$ \\
$D^{*+}\to \pi^+D^0\to\pi^+K^-\pi^+$ & mixing & 2.7 & 5000 & $10^7$ \\
$D^{*+}\to \pi^+D^0\to\pi^+K^-\pi^+\pi^+\pi^-$ & mixing & 5.5 & 3200 &
$0.6\times10^7$ \\
$D^{*+}\to \pi^+D^0\to\pi^+K^+\pi^-$ & DCSD & 0.02? & 45? & $10^4-10^5$ \\
$D^0\to K^-\pi^+\pi^+\pi^-$ & & 8.1 & & $6\times10^7$ \\
\hline
\end{tabular}
\end{center}
$^*$Note that the cuts
used (and hence the event yields) vary depending on the analysis goal,
thus the E791 yields display apparent inconsistencies at the factor-of-2 level.
\end{table*}

\end{document}